\documentclass{article}
\usepackage{spconf,amsmath,graphicx}
\usepackage[utf8]{inputenc}

\usepackage{algorithm}
\usepackage{algpseudocode}

\usepackage{hyperref}

\usepackage{tipa}

\usepackage{color}
\definecolor{todoColor}{rgb}{1,0,1}

\renewcommand{\paragraph}[1]{\noindent\textbf{#1}\quad}

\title{Repeat after me: Self-supervised learning of acoustic-to-articulatory mapping by vocal imitation}
\name{
    Marc-Antoine Georges$^{\star \dagger}$
    \quad Julien Diard$^{\dagger}$
    \quad Laurent Girin$^{\star}$
    \quad Jean-Luc Schwartz$^{\star}$
    \quad Thomas Hueber$^{\star}$
    \thanks{This work was funded by the Multidisciplinary Institute in Artificial Intelligence MIAI@Grenoble-Alpes (ANR-19-P3IA-0003).}
}
\address{
    $^{\star}$ Univ. Grenoble Alpes, CNRS, Grenoble INP, GIPSA-lab, 38000 Grenoble, France \\
    $^{\dagger}$ Univ. Grenoble Alpes, Univ. Savoie Mont Blanc, CNRS, LPNC, 38000 Grenoble, France
}

\begin{document}

\maketitle

\begin{abstract}
We propose a computational model of speech production combining a pre-trained neural articulatory synthesizer able to reproduce complex speech stimuli from a limited set of interpretable articulatory parameters, a DNN-based internal forward model predicting the sensory consequences of articulatory commands, and an internal inverse model based on a recurrent neural network recovering articulatory commands from the acoustic speech input. Both forward and inverse models are jointly trained in a self-supervised way from raw acoustic-only speech data from different speakers. The imitation simulations are evaluated objectively and subjectively and display quite encouraging performances.
\end{abstract}

\begin{keywords} Speech production, computational models, articulatory synthesis, representation learning.
\end{keywords}

\section{Introduction}

Speech production involves the activation of a complex system for uttering sounds from articulatory gestures. Learning to control such a complex system is a real challenge, considering the many underlying problems such as the large number of degrees of freedom of the vocal tract, the highly nonlinear relationship between articulatory configurations and sounds, the variability and complexity of the acoustical realizations available in the learning process and, most of all, the lack of corresponding articulatory information (apart from visual information on the facial movements of the interlocutor, if available). 
Therefore, from a statistical modeling point of view, learning the acoustic-to-articulatory relationship can be seen as a weakly supervised  process.

This hard problem has generated computational developments in basically three directions: (i) articulatory synthesizers enabling to generate vocal tract shapes and speech sounds from a restricted set of interpretable parameters \cite{maeda1990,birkholz2010model,birkholz2006construction}; 
(ii) statistical learning techniques modeling the articulatory-to-acoustic forward (e.g. \cite{hueber_2015_taslp_cgmr}) or inverse mapping (e.g. \cite{ bocquelet_2014_interspeech_articulatory_synthesis_dnn}) from in-vivo articulatory recordings; and
(iii) speech motor control architectures
mimicking how the human brain exploits internal sensory and motor representations in speech production. These architectures usually associate internal forward and inverse models \cite{miall1996forward}, respectively in charge of the articulatory-to-acoustic prediction and acoustic-to-motor-command inference (e.g. \cite{Guenther2012, parrell2019facts, parrell2019current}).
However, most of these models are tuned and tested on very simple data, such as vowels, simple syllables or synthetic data generated by the articulatory model itself (e.g. \cite{patri2015optimal,kroger2009towards, laurent2017, bailly1997learning,moulin-frier:14, philippsen2014learning, chen2021modeling}).

Another recent line of research focused on self-supervised learning of speech representations from large unlabeled audio datasets (e.g. \cite{chung19_interspeech}). These representations enable to train generative spoken language models that can be used to reproduce and extend a set of input audio stimuli \cite{polyak21_interspeech}.
A number of recent developments in this field occurred in the framework of the “zero-resource challenge” \cite{dunbar19_interspeech}.
However, the developed systems do not incorporate knowledge about the speech production process.
Therefore, although they are powerful and efficient for generating speech sounds, they cannot provide much light on the underlying mechanisms involved in speech production.

The present paper provides some progress at the crossroads between these two lines of research. %For this aim, we capitalize on %a speaker-specific articulatory model elaborated from a set of articulatory data and a classical architecture for developing articulatory models (Maeda 1990). This model has been shown to be articulatory realistic and able to accurately reproduce complex speech sequences (Georges et al., 2020). On this basis, %we propose a completely neuronal architecture including both a forward and an inverse model. We show that this system can indeed efficiently learn a corpus of natural speech from different speakers. We capitalize on an “accommodation” algorithm developed previously (Laurent et al., 2017) enabling the model to progressively shift from a random exploratory phase typical of infant babbling (ref) to a tuning phase in which the forward-inverse system converges to efficiently imitate the sounds of its human tutor.
We propose a fully neural architecture combining (i) a pre-trained  articulatory synthesizer (i.e., the plant) combining a deep neural network (DNN) based articulatory-to-acoustic mapping with the LPCNet neural vocoder  \cite{Valin2019} and able to reproduce complex speech stimuli from a limited set of interpretable articulatory parameters \cite{georges_issp2020,georges21_interspeech}, (ii) a DNN-based internal forward model predicting the sensory consequences of articulatory commands, and (iii) an internal inverse model based on a recurrent neural network (RNN) recovering articulatory commands from the acoustic speech input. Both the forward and inverse models are jointly trained in a self-supervised way from raw acoustic-only speech data from different speakers.  
%We show that this system can indeed efficiently learn a corpus of natural speech from different speakers. 
To that purpose, we investigate the “accommodation” algorithm proposed in \cite{laurent2017}, enabling the model to progressively shift from a random exploratory phase typical of infant babbling to a tuning phase in which the forward-inverse system converges to allow to efficiently imitate the sounds of a given speaker.

%In the following … PLAN.

\begin{figure}[ht!]
    \centering
    \includegraphics[width=1\linewidth]{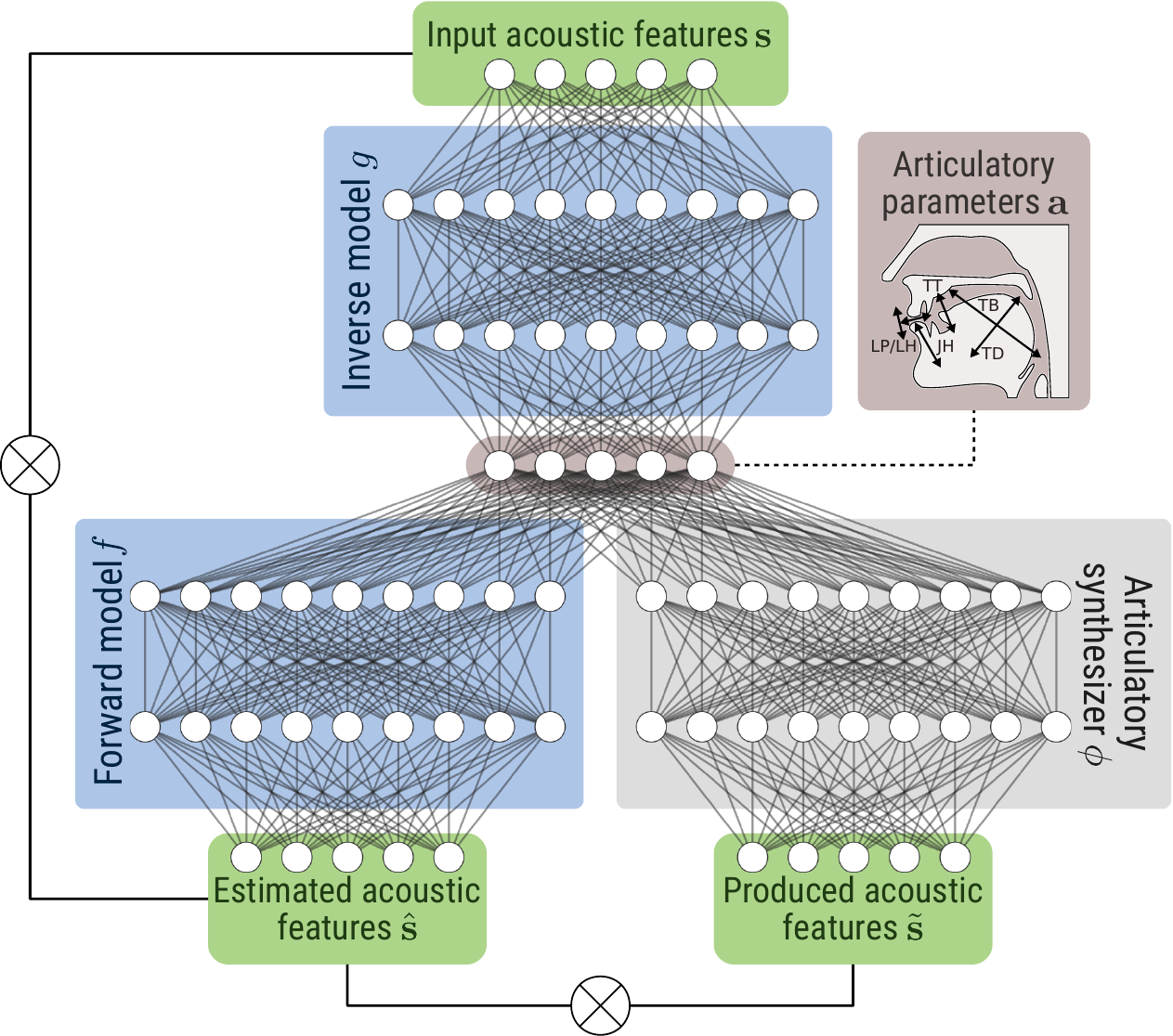}
    \caption{Overview of the proposed model of speech production.}
    \label{fig:architecture}
\end{figure}

\section{Material and methods}

\subsection{Overview}

%The proposed model represents a cognitive agent whose task is to learn how to repeat perceived speech.
%To do so, it has to find an articulatory sequence corresponding to its stimulus and use it to drive an articulatory synthesizer.
%As shown on figure \todo{ref}, the agent is composed of 3 components: the inverse model, the forward model and the articulatory synthesizer.
%The first two aim to model the cognitive part of the speech imitation process, whereas the last item models the physical part.
%During the learning phase, the inverse and forward models are trained together using an accommodation learning algorithm.
%The following sections describe the 3 components and the accommodation learning algorithm.

%First we set the notations used in the proposed model of speech motor control. Let use denote $\mathbf{s} = [\mathbf{s}_1,\cdots,\mathbf{s}_T]$ and  $\mathbf{a} = [\mathbf{a}_1,\cdots,\mathbf{a}_T]$  respectively a sequence of  $T$ acoustic features vectors and the corresponding vector of articulatory parameters (the nature of these vector $\mathbf{s}_t$ will be explicited in section \ref{sec:artisynth}). 

%Using its inverse model, the agent tries to find a set of articulatory commands $\mathbf{a}_{1:T}$ -- a sequence of articulatory parameters vectors -- that could produce the same sound.
%Then, using its forward model, it infers what would be the acoustic outcome $\hat{\mathbf{s}}_{1:T}$ of the execution of $\mathbf{a}_{1:T}$.
%Finally, the gesture $\mathbf{a}_{1:T}$ is produced using the articulatory synthesizer, resulting in $\Tilde{\mathbf{s}}_{1:T}$.

The proposed model is illustrated in Fig.~\ref{fig:architecture}. We briefly explain its expected behavior and set the mathematical notations.
Let us denote $\mathbf{s}$ an acoustic speech stimulus which is encoded into a sequence of $D_s$-dimensional feature vectors such as $\mathbf{s} = [\mathbf{s}_1,\cdots,\mathbf{s}_T]$ (where $T$ is the sequence length). The internal inverse model $g$ estimates at each timestep $t$ a set of articulatory parameters represented by the $D_a$-dimensional vector $\mathbf{a}_t = g(\mathbf{s}_t)$. At each timestep $t$, the internal forward model $f$ estimates the sensory (acoustic) consequence $\widehat{\mathbf{s}}_t=f(\mathbf{a}_t)$ of the articulatory input $\mathbf{a}_t$. Simultaneously, the articulatory synthesizer $\phi$ (i.e., the plant) simulates the physical process of speech production and generates the acoustic feature vector $\widetilde{\mathbf{s}}_t = \phi(\mathbf{a}_t)$.  The precise nature of the acoustic ($\mathbf{s}_t$, $\widehat{\mathbf{s}}_t$ and $\widetilde{\mathbf{s}}_t$) and  articulatory features ($\mathbf{a}_t$) will be detailed in Section~\ref{sec:artisynth}. Here, $g$, $f$ and $\phi$ are DNNs. 

When trained to reproduce an input audio speech stimulus (i.e., minimizing the discrepancy between $\mathbf{s}_t$ and $\widehat{\mathbf{s}}_t$ at each timestep), this model learns a forward and inverse articulatory-to-acoustic mapping in a self-supervised manner. 
The following sections describe the three components in more details, as well as the accommodation learning algorithm used to jointly train the internal models from audio-only speech stimuli.

\subsection{Neural articulatory synthesizer}
\label{sec:artisynth} 
In the present study, we use the neural articulatory synthesizer originally proposed and described in \cite{georges_issp2020} (and recently used in \cite{georges21_interspeech}). We recall here the key features of this synthesizer. %(see Fig.~\ref{fig:articulatory_synthesizer}). 

First, it was built from  an articulatory-acoustic dataset containing recordings of 1,108 items (vowels, vowel-consonant-vowel sequences, short words and sentences) pronounced by a male French speaker (hereafter the reference speaker RS). The movements of the tongue, jaw and lips during speech production were recorded at 100~Hz using 2D electromagnetic articulography (EMA), synchronously with the speech sound. An articulatory model accounting for the different degrees of freedom of the vocal tract was then built from the raw EMA coordinates using the guided-PCA method proposed in \cite{maeda1990}. Using this model, EMA coordinates  were then converted into a limited number of interpretable articulatory parameters: \textit{Jaw Height} (\textit{JH}),
\textit{Tongue Body} (\textit{TB}),
\textit{Tongue Dorsum} (\textit{TD}),
\textit{Tongue Tip} (\textit{TT}),
\textit{Lip Protusion} (\textit{LP}), and
\textit{Lip Height} (\textit{LH}).
In parallel, the short-term spectral content of the acoustic speech signal (initially recorded at 16~kHz) is encoded into a vector of 18 Bark-scale cepstral coefficients, which is the acoustic feature vector $\mathbf{s}_t$ (the analysis is processed with a window size of 20~ms and a hop size of 10~ms, ensuring the synchronization of the acoustic and articulatory feature vectors). 

The articulatory-acoustic relationships for speaker RS were then approximated by a feedforward DNN (4 fully-connected hidden layers of 256 neurons each, with an hyperbolic tangent activation function, batch normalization and dropout ($p=0.25$) used in each layer, trained on 80\% of the dataset with the Adam optimizer and mini-batches of 8 observations, the remaining 20\% being used for early-stopping).
At synthesis time, a synthetic speech waveform can be obtained using the LPCNet neural vocoder, conditioned on the  speech spectrum estimated from the articulatory parameters, and combined with source features (i.e., $f_0$ and harmonicity parameters in LPCNet). 

Importantly, this neural articulatory synthesizer is used here as a pre-trained module. When used within the proposed architecture, its parameters are not updated contrary to the inverse and forward models $g$ and $f$, which are described in the next section.  

\subsection{Implementation of the forward and inverse models}

The forward model $f$ (converting an articulatory feature vector $\mathbf{a}_t$ into an acoustic feature vector $\widehat{\mathbf{s}}_t$) is implemented with the same architecture as that of the articulatory synthesizer (4 fully-connected hidden layers of 256 neurons each, trained with the same settings as those reported in Section~\ref{sec:artisynth}). 

The inverse model $g$ in charge of recovering the articulatory parameters that must be sent to the plant $\phi$ to reproduce the input speech stimuli is implemented with an RNN. This choice is motivated by the one-to-many nature of the acoustic-to-articulatory mapping, 
which may be partially tackled by considering contextual information. %represents the process of inferring the articulatory gestures that could be responsible of a speech sound.
%When the agent needs to reproduce a sound, it uses his inverse model to find the adequate sequence of articulatory parameters.
%Like the forward model, the inverse model starts with no knowledge and improves during the learning process.
%Due to the one to many nature of the acoustic-to-articulatory mapping, we implemented the inverse model as a Long Short-Term Memory (LSTM).
%The temporal nature of the LSTM allows it to draw contextual information when finding the articulatory command behind a certain frame of an acoustic sequence.
%In this work the inverse model consists of 
Therefore, we use 2 layers of 32 long short-term memory (LSTM) cells (with dropout $p=0.25$), stacked with a linear (time-distributed) layer, for converting a sequence of $T$ acoustic feature vectors $\mathbf{s}=[\mathbf{s}_1, \cdots, \mathbf{s}_T]$ into a sequence of $T$ articulatory feature vectors $\mathbf{a}=[\mathbf{a}_1, \cdots, \mathbf{a}_T]$. 

%Each layer has an hidden state of 32 features, and are trained with dropout ($p=0.25)$.
%A linear layer is used to project the output of the last layer to the articulatory parameters space.

% - Is a temporal model due to the one to many nature of the problem
%   - Pour aider, we rely on contextual information

\subsection{Accommodation learning}

% Accomodation learning is defined here as a learning process, during which the model tries to imitate external sound stimuli, all the while gathering observations about gestures and sounds (both the initial target sounds and, crucially, the ``real'' sounds generated by gestures).
Accomodation learning is defined here as a learning process, during which an agent tries to imitate external sound stimuli by generating articulatory commands, and gathers observations about the correspondence between target sounds and attempted articulations on one hand, attempted articulations and real sounds generated by these articulations on the other hand.
Starting from a random initialization, the procedure used to jointly train the forward and inverse models can be summarized as follows: (i) given a sequence  of acoustic observations $\mathbf{s}$ (in practice, a mini-batch of $K$ sequences with $K = 8$ in our implementation), estimate $\mathbf{a}=g(\mathbf{s})$ (inverse model), $\widehat {\mathbf{s}}$ with $\widehat {\mathbf{s}}_t=f(\mathbf{a}_t)$ (forward model) and $\widetilde{\mathbf{s}}$ with $\widetilde{\mathbf{s}}_t=\phi(\mathbf{a}_t)$ (articulatory synthesizer) (ii) update the forward model $f$ using backpropagation to minimize the discrepancy between $\widehat{\mathbf{s}}$ and $\widetilde{\mathbf{s}}$ (in practice, the mean squared error is used as the loss function), and (iii) update the inverse model $g$ to minimize the discrepancy between $\mathbf{s}$ and $\widehat{\mathbf{s}}$, while keeping the forward model unchanged. This is achieved by freezing the parameters (i.e., the weights and biases) of the forward model during the backpropagation of the prediction error from the last layer of the forward model to the first layer of the inverse model). %This algorithm is detailed in pseudo-code in Alg. \ref{?}. 

When starting the accomodation learning, no prior knowledge is present in the inverse and forward models. Therefore, the inversion step yields unadapted productions, which amounts to randomly explore the articulatory space. However, applying such random gestures as inputs to $\phi$ provides ``ground-truth'' acoustic consequences, allowing $f$ to locally learn the articulatory-to-acoustic mapping. Over time, learning progresses and focuses on relevant regions of the articulatory space, with an improvement of the forward model leading to a better inversion, improvement of the inverse model and of the articulatory gesture selection, and finally, of the imitation task.

\begin{figure}[ht!]
    \centering
    \includegraphics[width=0.95\linewidth]{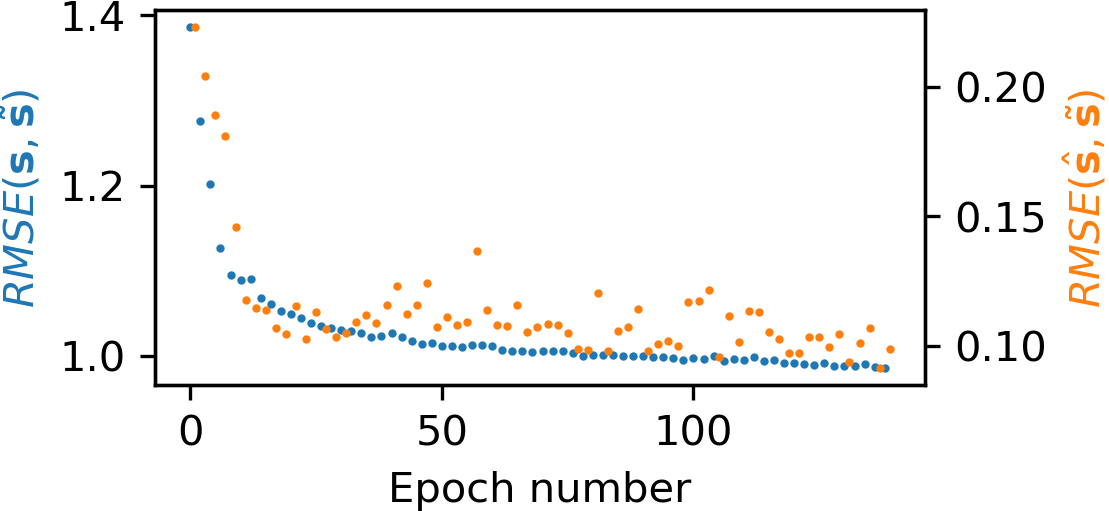}
    \vskip -0.3cm
    \caption{RMSE evolution for the forward (in orange) and inverse (in blue) models ($y$-axis) as a function of the number of epochs ($x$-axis), for a simulation in which RS imitates S1.}
    \label{fig:loss_progress}
\end{figure}

\begin{figure}[ht!]
    \centering
    \includegraphics[width=1\linewidth]{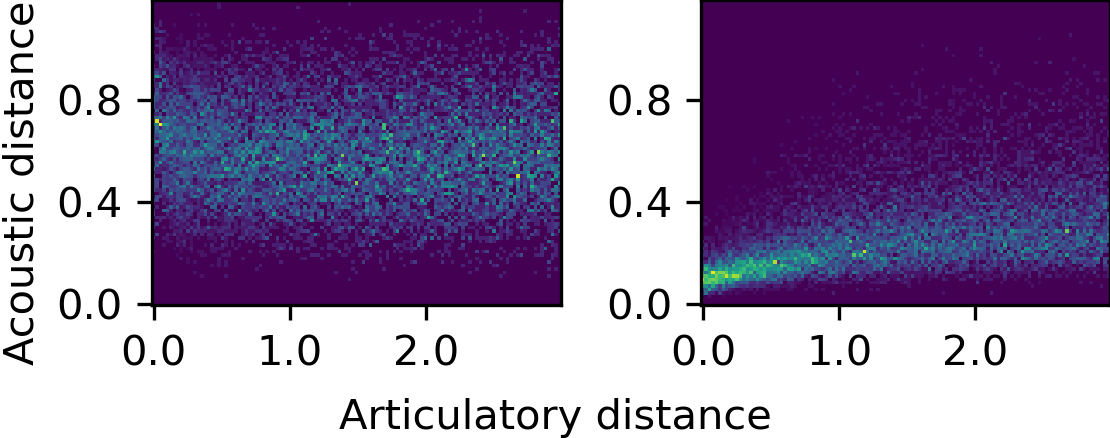}
    \vskip -0.3cm
    \caption{Accuracy of the forward model ($y$-axis) as a function of the distance between an articulatory input and the articulatory set selected at the last epoch ($x$-axis). Left: at learning onset; Right: at the end of learning.}
    \label{fig:direct_model}
\end{figure}
\section{Simulations}

\subsection{Datasets}

Model training and evaluation were conducted on 3 audio-only datasets. The first one is derived from the audio recordings of the reference speaker (RS) used to train the articulatory synthesizer (see Section \ref{sec:artisynth}). In that case, the model perceives (and tries to repeat) its own ``voice." The two other audio datasets were built by recording two other male speakers (hereafter S1 and S2) uttering the same set of speech utterances.

For each of the 3 speakers considered in this study (RS, S1, S2), 80\% of the items in the dataset were selected randomly and used for training while the remaining 20\% were used for test.  Among the training items, 20\% of them were used for validation (early stopping). For each speaker, we repeated the simulation 3 times. The experimental results are reported and discussed in the following sections. 

\subsection{Learning dynamics}

%\paragraph{Learning curves}
Fig. \ref{fig:loss_progress} displays the typical evolution of the RMSE values for the forward and inverse models (that is, between $\widehat{\mathbf{s}}$ and $\widetilde{\mathbf{s}}$ for the forward model $f$ and between $\mathbf{s}$ and $\widetilde{\mathbf{s}}$ for the inverse model $g$) along the model training for one simulation. Learning is rapid, though less smooth for $f$ than for $g$, since it is trained on the productions inferred by the constantly evolving inverse model, causing instabilities. RMSE at convergence is quite low for $f$ since it is progressively tested on specific regions of the articulatory-acoustic space where it is particularly well trained and accurate.

%\paragraph{Evaluation of the forward model}
Fig. \ref{fig:direct_model} further evaluates the accuracy of the forward model at the beginning and at the end of the learning phase. For this aim, we first gathered the whole set of articulatory feature vectors generated by the model when processing the test set at the end of the learning process. These vectors were then randomly modified with an offset and fed into the forward model and the articulatory synthesizer. We then computed the difference between the resulting outputs.
The right plot of Fig. \ref{fig:direct_model} displays this difference sorted by the distance between the articulatory feature vector selected by the model and its randomly offset version. The left plot of the figure displays the same process on the forward model at the beginning of the learning phase (Epoch 1).
The results show that at learning onset, the forward model is consistently bad at inferring acoustic outcomes.
In contrast, after learning, the forward model is accurate for articulatory feature vectors close to the ones the inverse model has selected, while accuracy decreases further away from these selected vectors. This shows that the accommodation learning algorithm only learns the forward model in small, useful portions of the articulatory space.

\subsection{Evaluation of imitation performance}

The models were then tested on the quality of the phonemic content of the produced speech signal, with both an objective evaluation by an HMM-based phonetic decoder and a subjective evaluation by human listeners.

\begin{figure}[htb]
    \centering
    \includegraphics[width=0.92\linewidth]{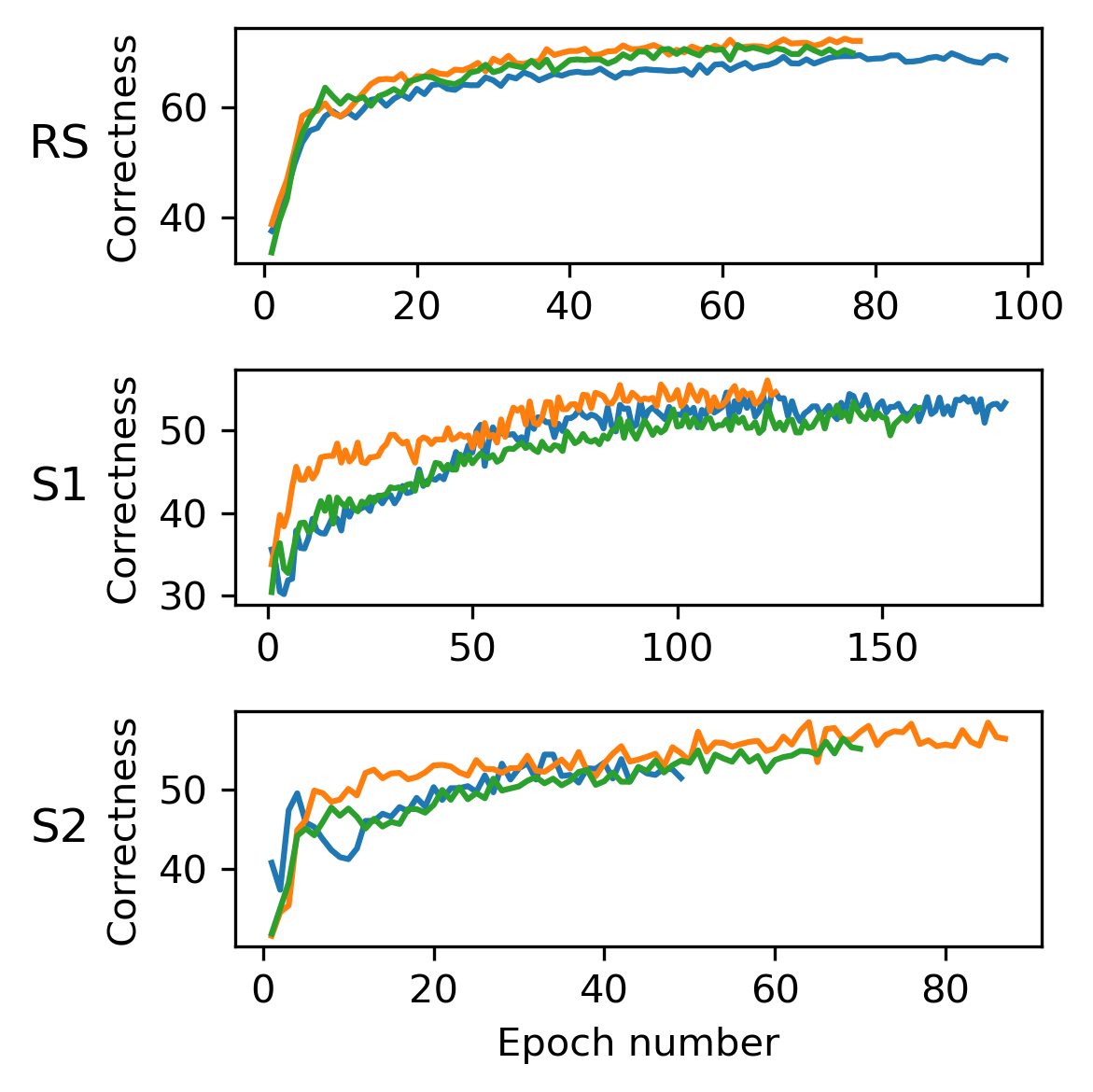}
    \vskip -0.25cm
    \caption{Evolution of the correctness (in percent, $y$-axis) of the HMM-based phonetic decoding of the model productions as a function of the number of model training epochs ($x$-axis). Each of the 9 curves represents a different simulation trained to reproduce one dataset.}
    \label{fig:reco_hmm}
\end{figure}

%\paragraph{Objective evaluation} 
A phonetic decoder based on a set of 39 3-state left-to-right HMMs  was first trained on the audio dataset of speaker RS, using the same items and acoustic features as those used to train the forward and inverse models (training was done using the HTK toolkit and a standard procedure). Then, at each learning epoch, the test set outputs of the 9 simulations were fed into the HMM system providing correctness scores at the phonetic level, which are displayed on Fig. \ref{fig:reco_hmm}.
These scores have similar learning dynamics as those of RMSE values of the forward and inverse learning models in Fig. \ref{fig:loss_progress}. Correctness reaches values around 70\% for the imitation of RS, and lower values, around 60\%, for the two other speakers. This is not surprising considering that the content of the phonetic utterances produced by speakers S1 and S2 differ from the productions by speaker RS, which provided the basis for the HMM training. Still, the global quality of the produced utterances at the end of the training stage is rather satisfactory.\footnote{Productions associated to these simulations are available at \url{https://georges.ma/p/imitative-model}.} 

%\paragraph{Subjective evaluation}
To better evaluate the quality of the imitation test set for each imitated speaker, we extracted from the test corpus a set of isolated oral vowels /\textipa{i}  \textipa{y} \textipa{u}  \textipa{e} \textipa{\o} \textipa{o} \textipa{a}/ and a set of non-nasal consonants /\textipa{b}  \textipa{d} \textipa{g} \textipa{p}  \textipa{t} \textipa{k} \textipa{f}  \textipa{v} \textipa{s}  \textipa{z} \textipa{S}  \textipa{Z}/ in a VCV context, with V one of the three vowels /\textipa{i}  \textipa{a} \textipa{u}/. This set of stimuli, separately produced by each of the three speakers, was presented to the model for imitation, carefully ensuring that the corresponding model had not been trained on the stimulus to imitate. The resulting 18-Bark band spectrum was converted into sound by using the LPCNet vocoder with the source parameters from the original sound to imitate. The set of synthesized stimuli for each of the three imitated speakers was presented to 20 listeners in an online test in the Prolific platform, with a forced choice response. %The mean correct scores for consonants and for vowels for each of the imitated speaker are provided in Table 1. 
The average correct scores for consonants are around 60\% (58 \% for RS, 62 \% for both S1 and S2). For vowels, they vary from 94\% from speaker RS to lower scores, respectively 54\% and 72\% for S1 and S2. The lower scores for speakers S1 and S2 are probably due to some inconsistency between vowel targets among the target speakers, confusions arising typically between close vowels in the formant space (e.g., /\textipa{y}/ vs. /\textipa{e}/).

% \begin{figure}[htb]
%     \begin{center}
%         \begin{tabular}{ |c|c|c|c| } 
%         \hline
%         Target speaker & Vowels & VCV \\
%         \hline
%         RS & 94.05\% & 58.48\% \\ 
%         S1 & 71.67\% & 61.90\% \\ 
%         S2 & 53.57\% & 61.61\% \\
%         \hline
%         \end{tabular}
%     \end{center}
%     \caption{Experimental results of the perception test.}
%     \label{fig:perceptive_test}
% \end{figure}

\section{Conclusion and perspectives}
The presented simulations  show that it is indeed possible to train in a weakly supervised way a sensory-motor model to jointly learn a forward and an inverse model from speech samples provided by different speakers. The end-to-end neuronal architecture presented and evaluated in the present paper provides powerful ways of adapting sensory-motor architectures to the real speech world. Still there is much room for improvement of the proposed architecture, particularly concerning its ability to better define the speech task in terms of phonetic and linguistic goals rather than strict imitation of acoustic patterns. The next step in these developments will hence consist in introducing linguistic units in the definition of representations and loss functions.

\bibliographystyle{IEEEbib}
\bibliography{refs}

\end{document}